# Analysis of Digitalized ECG Signals Based on Artificial Intelligence and Spectral Analysis Methods Specialized in ARVC

**Vasileios E. Papageorgiou[1,*], Thomas Zegkos[2], Georgios Efthimiadis[2] and George Tsaklidis[1,*]**

[1]Department of Mathematics, Aristotle University of Thessaloniki, Thessaloniki 54124, Greece
[2]1st Cardiology Department, AHEPA University Hospital, Aristotle University of Thessaloniki, 54636, Thessaloniki, Greece

**\*Correspondence:** tsaklidi@math.auth.gr; vpapageor@math.auth.gr

## Abstract

Arrhythmogenic right ventricular cardiomyopathy (ARVC) is an inherited heart muscle disease that appears between the second and forth decade of a patient's life, being responsible for 20% of sudden cardiac deaths before the age of 35. The effective and punctual diagnosis of this disease based on Electrocardiograms (ECGs) could have a vital role in reducing premature cardiovascular mortality. In our analysis, we firstly outline the digitalization process of paper – based ECG signals enhanced by a spatial filter aiming to eliminate dark regions in the dataset's images that do not correspond to ECG waveform, producing undesirable noise. Next, we propose the utilization of a low – complexity convolutional neural network for the detection of an arrhythmogenic heart disease, that has not been studied through the usage of deep learning methodology to date, achieving high classification accuracy, namely 99.98% training and 98.6% testing accuracy, on a disease the major identification criterion of which are infinitesimal millivolt variations in the ECG's morphology, in contrast with other arrhythmogenic abnormalities. Finally, by performing spectral analysis we investigate significant differentiations in the field of frequencies between normal ECGs and ECGs corresponding to patients suffering from ARVC. In 16 out of the 18 frequencies where we encounter statistically significant differentiations, the normal ECGs are characterized by greater normalized amplitudes compared to the abnormal ones. The overall research carried out in this article highlights the importance of integrating mathematical methods into the examination and effective diagnosis of various diseases, aiming to a substantial contribution to their successful treatment.

## 1. Introduction

Arrhythmogenic right ventricular cardiomyopathy (ARVC), is an inherited heart muscle disease characterized by fibro-fatty replacement of the right ventricular myocardium that predisposes patients to arrhythmia and right ventricular (RV) dysfunction leading in some cases to sudden cardiac death (SCD) [1 – 3]. It has been reported that this heart disease has prevalence of 1 in 2000 to 5000 individuals; however, many studies demonstrated that this rate is an underestimation of the real disease's prevalence and in certain geographic regions it may reach a rate of 1 in 1000 [4]. Importantly ARVC is responsible for 20% of SCDs before the age of 35 [5 – 6]. ARVC is inherited with the autosomal dominant pattern with incomplete, however, penetrance and variable expressivity [7], while the onset of the disease occurs between the second and the fourth decade of patients' life. Nevertheless, a SCD may occur early in adolescence, without warning symptoms or structural abnormalities that could lead to timely diagnosis [2]. Moreover, the association between ARVC and sport activities is well-known, as a significant percentage of SCDs concerning athletes have been attributed to this heart muscle disease. To lend more support, intense physical exercise can lead to disease progression [8].

One of the most widely used tools for ARVC diagnosis are cardiovascular magnetic resonance (CMRI) and echocardiography. During the last decade, CMRI represents a useful medical tool for the efficient diagnosis of ARVC, permitting non-invasive morphological and functional evaluation accompanied by tissue characterization [9]. The most important and easy diagnostic exam for the diagnosis of ARVC, however, is the electrocardiogram (ECG). In general, various diagnostic criteria for the existence of ARVC can be assembled by the ECG assessment [10 – 14]. ECG criteria are usually evaluated in combination with the criteria of echocardiography and CMRI.

More specifically, the ECG constitutes a valuable diagnostic test for patients suffering from ARVC, due to the recorded repolarization and/or depolarization abnormalities in about 90% of cases [15]. The set of diagnostic criteria is divided into two smaller categories, the categories of major and minor criteria with respective diagnostic value. The major ECG criterion for ARVC, is the presence of epsilon waves that are most prominent in the right precordial leads (V1 – V3) [16]. Epsilon waves were first reported in 1977, and represent a low-amplitude signal, between the end of QRS complex and the beginning of T wave, that can be seen in patients with advanced or late ARVC stage. These abnormal waves are due to islands of the surviving cardiomyocytes that are restricted or embedded in interstitial fibrosis or fat, a phenomenon that causes delayed activation of the right ventricle, which results in delayed potentials [17].

There are three minor criteria that characterize the existence of ARVC: low QRS voltages ($< 0.5$mV) in the limb leads, QRS delayed S-wave upstroke with terminal activation duration (TAD) $\geq$ 55ms in the right precordial leads and negative T waves usually observed in V1 and V2 leads [18]. The QRS delayed S-wave upstrokes are often accompanied by negative T waves, while the presence of negative T waves in V1 and V2 leads may be considered as normal in cases where the individual does not meet additional diagnostic criteria [15]. Generally, the presence of a minor criterion individually, does not imply the existence of the disease in the subject.

In many articles elements of abnormal ECG signals that correspond to patients diagnosed with ARVC are examined, and their differentiation from ECGs belonging to healthy individuals is investigated. Indicatively, in [19] the authors compare the characteristics of P waves in patients with ARVC, 10 years before and up to 15 years after the diagnosis of the disease, concluding that initial ARVC progression is accompanied by P wave flattening while later stages are characterized by increased prevalence of PTF-V1. Moreover, the authors in [20] examine the role of premature ventricular ablation in patients with ARVC, while [21] describes the symptoms and abnormalities encountered in women suffering from this disease. Finally, in [22] the prevalence of left ventricular (LV) involvement and imaging features of LV phenotype in patients with ARVC is highlighted, while the authors in [23] investigate the existence of disease-associated microRNAs in pericardial fluid that are potentially linked to ARVC pathogenesis.

The analysis displayed in our paper, could be separated into three parts; the digitalization of paper-based ECG signals, the creation of an automatic classification process of normal ECG heartbeats and heartbeats corresponding to patients suffering from ARVC using a convolutional neural network (CNN), and the investigation of spectral differentiations between the aforementioned ECG categories.

The first part includes a brief presentation of the digitalization process used for the exploitation of paper-based ECG signals. This preprocessing step has a vital role in our research, due to the requirement for digitalized time series, aiming to compare the amplitude spectrums of normal and abnormal ECGs. The authors in [24 – 26] display an ECG digitalization process targeting the elimination of signal's noise. During the digitalization process, we propose an improved version of the filtering process proposed in [27] that eliminates isolated black pixels that do not represent the ECG waveform. Our algorithm eliminates black pixels that do not correspond to the signal by examining the values of neighboring pixels and not just the horizontally previous and the next pixel, as it is presented in section 2.

The part of ECG classification proposes the usage of a low – complexity CNN that classifies heartbeats into normal and abnormal ARVC beats. Furthermore, we take advantage of the Bayes theorem to calculate the exact probability of a patient suffering from ARVC by examining multiple heartbeats. Many articles present the utilization of a CNN to classify arrhythmias, although none of them includes a class for individuals suffering from ARVC. An 1D – CNN is utilized to classify normal and abnormal ECG heartbeats into 3 classes [28 – 29], while in [30 – 31] the classification concerns one and two additional types of arrhythmias respectively. Other articles highlight the usage of 2D – CNNs for multiclass heartbeat classification, e.g. [32] proposes a 11 – layer 2D – CNN to classify ECGs belonging to MIT-BIH arrhythmia database into normal, supraventricular, ventricular, fusion and unknown class with 98.3% accuracy, while a 5 class classification of arrhythmias based on a 2D – CNN is also presented in [33 – 34]. Furthermore, the authors in [35] examine the diagnostic capacity of another 11 – layer CNN classifying normal and abnormal heartbeats into 10 classes, while in [36 – 37] the diagnostic accuracy of multilayer CNNs is investigated on different datasets consisting of 5 and more classes. Another group of articles [38 – 40], proposes the utilization of CNNs for multiclass classification, with the particularity that the inputs of these deep learning algorithms are not the waveforms of heartbeats themselves, but the spectrograms produced by the respective ECG signals. A 3D–CNN, namely AlexNet, is described in [41] for the classification of normal, right bundle brunch block and paced beats, whereas in [42 – 43] the equivalent process is relied on the combination of CNN–LSTM models, where LSTMs belong to the architecture class of the recurrent neural networks (RNN). A review of the utilized deep learning approaches that concern ECG arrhythmia classification is presented in [44].

In the third and final part, the significant elements of the amplitude spectrums of normal and abnormal ECGs are compared by testing the significance of these differences via parametric and non-parametric statistical tests. For the construction of amplitude spectrums, we used the fast Fourier Transform (FFT); denoising is one of the most common applications of FFT in signal analysis, e.g., in [45] it is presented the usage of an IIR filter for the denoising of ECG signals. The authors in [46] describe a simulation methodology for the production of ECG waveforms based on power spectral analysis, in [47] the finite number of necessary frequencies that are needed for the efficient reconstruction of a normal ECG signal is investigated, while in [48] useful waveform elements are extracted from ECGs through FFT that are classified by a multi-objective genetic algorithm. There are many articles that compare the produced ECG spectrums between various categories of individuals, e.g., [49] performs spectral analysis concerning 6 normal adults and 8 normal children, while in [50] the power spectrums of 3 groups containing 32 post myocardial infraction patients with ventricular tachycardia (VT), 19 without VT and 17 healthy individuals, are investigated and compared. The authors in [51] compare the power spectrums between control subjects and patients that are cardiac transplant recipients; in [52] a power spectral analysis is presented utilizing ECGs belonging to individuals suffering from diabetes, while in [53] the normalized amplitude spectrums of normal beats and beats displaying 5 arrhythmias are compared. In [54] the utilization of power spectral analysis in more advanced stages of congestive heart failure allows the identification of a subgroup of subjects with major risk of adverse events, while in [55] spectral analysis of ECG segments representing 16 participants with and without the consumption of anhydrous glucose solution is investigated.

Our analysis provides a complete methodology for the digitalization, the classification of normal and abnormal ECG signals and the examination of their spectral characteristics in contrast to various other studies emphasizing on only one of these aspects. Moreover, the proposed spatial filter is adequate to effectively eliminate dark regions in the ECG images that could highly distort the digitalization results, enabling us to even take advantage of paper-based ECGs of low quality. The classification accuracy of the proposed CNN, highlights the usefulness of artificial intelligence (AI) methods that are trained based on the knowledge of experienced doctors who can rely their diagnosis on additional diagnostic criteria. Undoubtedly, the ECG criteria that are considered for the diagnosis of ARVC, constitute a very important tool, but they lack specificity. ECG is abnormal at around 90% of ARVC patients, however, none of the reported abnormalities can safely distinguish ARVC from other cardiac conditions [14]. Therefore, other task force criteria have to be utilized to ensure the diagnosis. On the other hand, CNN provides significantly higher diagnostic accuracy (98.6% for test data) based solely on ECG, as its training has been applied to ECG instances considered as abnormal, with the aid of other supplementary medical tools and not only by their visual examination.

Compared to other studies that examine only one or two ECG characteristics like the morphology of T or the epsilon waves to base their diagnosis, our method includes in the classification procedure all major and minor diagnostic criteria as it utilizes the complete ECG beat and not only individual criteria. In parallel, our methodology proposes a low – complexity CNN ideal for small datasets, that usually accompany medical studies compared to aforementioned articles which utilize more complex and computationally expensive architectures. This low – complexity structure not only eliminates the emergence of overfitting, but also offers flexibility as it can be easily retrained in case new ECGs are added to the dataset. Finally, the classification and the spectral analysis parts, aim to reveal pattern differentiations between normal and ARVC ECGs providing new perspectives for the examination of this difficult to identify cardiovascular disease, while this is the first time that it is explored using mathematical modeling techniques.

This article is organized as follows: In section 2 we outline the utilized methods and mathematical tools for the digitalization and CNN classification process as long as the respective tools for the third part of spectral analysis. In section 3, we display the results that are produced by applying the aforementioned methods on normal ECGs and ECGs from patients who suffer from ARVC. Section 4 provides a discussion concerning the results presented in the previous section, while section 5 highlights useful conclusions that summarize the proposed analysis, accompanied by some points of interest in the medical field, which are derived from the present research.

## 2. Methods and Mathematical tools

In this section, we introduce the mathematical models and methods used in the three parts of our analysis. The first part of this section concerns the synoptic description of the digitalization process utilized to produce a usable ECG signal version on which our mathematical analysis is relied. Moreover, we propose an artificial intelligence model, which is suitable for classifying normal and pathological ECG heartbeats. Our analysis is completed with the presentation of some elements concerning the field of frequency and spectral analysis.

*2.1. Digitalization of Paper – Based ECG Records.*

A common problem during medical research and more specifically during ECG signal analysis, is that every hospital has a wide dataset of paper – based ECGs that are sufficient only for visual examination. A reliable digitalization method would solve this problem paving the way for more systematic research using various mathematical tools. We will describe the process of converting paper-based ECG data to a 1-D digitalized signal.

2.1.1. Conversion from scanned to binary image.

Our approach begins with the scanning of the paper–based ECG to produce a digital image that is converted to 8-bit grayscale with normalized values. Our grayscale image is rendered with a two-dimensional matrix $\mathbf{G} = (G_{i,j})$ containing real values from 0 to 1 (black = 0 and white = 1) for each of its pixels. Every ECG paper contains the ECG signal, which is printed on a graphical grid that aids the measurement of useful time intervals and magnitudes. As a result, we need a thresholding technique that will eliminate graphical grid without distorting the signal, aiming to create a binary image. A reliable threshold can be obtained using the histogram of values of the above matrix. Provided that the graphical grid is usually colored with lighter colors while the part corresponding to the ECG signal is black, the majority of the values of the pixels approach 1. According to the above observation, we define the threshold to equal the 95th percentile of pixels distribution $F(x)$ [25]. Hence, the binary image is described by a matrix $\mathbf{BW} = (BW_{ij})$, with only ones and zeros that are derived from the conditions

$$G_{ij} > x_{0.95} \Rightarrow BW_{ij} = 1 \tag{1}$$

and

$$G_{ij} \leq x_{0.95} \Rightarrow BW_{ij} = 0, \tag{2}$$

where

$$F(x_p - 0) \leq p \leq F(x_p). \tag{3}$$

and for our case $p$ equals 0.95.

On this binary image, a column-wise pixel scan is performed to find the contour of the ECG signal based on the location of black pixels. Furthermore, each column may have more than one zero value (black pixel) because of the thickness of the ECG waveform. As a result, we obtain a pair of time-voltage values for each point of the mean contour of the ECG waveform, where the mean contour is constructed by the middle black pixels of each column. The sampling rate of our signals depends on the resolution of our scanned images that is 600 dpi.

2.1.2. Filtering incorrectly classified black pixels.

Based on the previous process, an important problem emerges derived from a wide collection of ECG papers that hospitals own, printed from a wide variety of existing cardiographs. Our dataset contains instances of ECG papers with darker graphical grid or with various abnormalities in its colors, leading to incorrectly classification of black pixels that they do not belong to the ECG waveform degrading the robustness of the aforementioned method.

Our approach involves the usage of a k × k local mean filter, denoted as $\boldsymbol{M}$, which is a k × k matrix that will automatically transform incorrectly black pixels to white pixels, by checking the allocation of black pixels in the subregions of the binary matrix $\boldsymbol{BW}$. Our algorithm gives more emphasis to the outer layer, reinforcing the filter's effectiveness in cases of small regions containing closely coiled incorrect black pixels. The filtering takes place by scanning the n × m binary matrix. When

$$BW_{ij} = 0, \forall i \in \left\{ \left\lceil \frac{k}{2} \right\rceil, n - \left\lfloor \frac{k}{2} \right\rfloor \right\}, \qquad j \in \left\{ \left\lceil \frac{k}{2} \right\rceil, m - \left\lfloor \frac{k}{2} \right\rfloor \right\}, \tag{4}$$

then, if

$$\frac{1}{k^2 - 1} \sum_{i - \lfloor k/2 \rfloor}^{i + \lfloor k/2 \rfloor} \sum_{j - \lfloor k/2 \rfloor}^{j + \lfloor k/2 \rfloor} BW_{lq} \times m_{lq} > c, \tag{5}$$

for some c ∈ ℝ⁺, this element is converted to $BW_{ij} = 1$. Moreover, we define

$$BW_{1j} = \cdots = BW_{\left\lfloor \frac{k}{2} \right\rfloor j} = BW_{n - \left\lfloor \frac{k}{2} \right\rfloor + 1 j} = \cdots = BW_{nj} = 1, \quad \forall j \in \{1, \dots, m\}, \tag{6}$$

and

$$BW_{i1} = \cdots = BW_{i \left\lfloor \frac{k}{2} \right\rfloor} = BW_{i m - \left\lfloor \frac{k}{2} \right\rfloor + 1} = \cdots = BW_{im} = 1, \quad \forall i \in \{1, \dots, n\}. \tag{7}$$

After experimentation, we propose the usage of a 5 × 5 mean filter while $c = 0.5$. The 5 × 5 spatial extent provided better results than the 3 × 3 filter, while for k > 5, the computational complexity of the algorithm grows without remarkable enhancements in the filter's performance. We chose $k^2 - 1$ as the denominator of the aforementioned sum, because we already know that the central element of the examined region will always equal 0, as we are only interested in incorrect black pixels. Simultaneously, we suggest a weighted sum by defining the sum of weights of the outer layer of the filter to equal 0.6 and the sum of the inner layer as 0.4, while the value of the central element of the filter can be any real number, provided that the central pixel will always be a black pixel of value 0. Hence, the correction of incorrectly classified black pixels depends on the pixels of their vicinity. Apparently, as c → 0 more and more black pixels will be transformed into white ones.

## 2.2. Signal Preprocessing.

The digitalization process of the signal is the first preliminary step to acquire a usable ECG signal. The methods we will describe below correct two elementary problems that usually distort the quality of an ECG signal, aiming to enhance the reliability of the following analysis.

2.2.1. Baseline Wandering Correction.

After the digitalization of the paper-based ECG signal, we usually face the problem of local trends integrated in the signal. These trends are produced during the printing of the ECG signal on the paper and most of the times are deemed as failed cardiograph attempts to display the signal correctly. The key element of this phenomenon is the occurrence of trends on the isoelectric line (ISO), which is the part of ECG connecting T and P waves and should be around 0 mV [24].

These trends can be eliminated using a moving median filter of order 121, obtained after experimentation on our dataset. Let $\{x_t\}$ be the timeseries of the ECG signal, consisted of $n'$ observations. For each $x_i$ we take the median value $z_i$ of a time window of 121 observations (60 time points before and 60 after $x_i$) and construct a timeseries of these values denoted as $\{z_t\}$, where

$$z_t = median(x_{i-60}, x_{i-59}, \dots, x_{i+59}, x_{i+60}), \tag{8}$$

with $t = 1, \dots, n'$ while $n'$ denotes the length of the time series $\{x_t\}$. We overcome the problem of edge effects using reflection, maintaining the length of the initial timeseries. The new detrended timeseries $y_t = x_t - z_t$, contain all the spatial information of the initial signal while ISO line is now around 0.

2.2.2. Smoothing.

After correcting the position of the ISO line, we can optionally smooth the produced digitalized signal using Savitzky-Golay filter to reduce noise added to the signal from the cardiograph through the printing process. This smoothing procedure operates as a low-pass filter, eliminating the high frequency elements of the digitalized ECG signal. It is a digital filter applied to increase the precision of the data without distorting the signal morphology, which is achieved through convolution, by fitting a low degree polynomial to successive windows of time points based on the method of linear least squares.

*2.3. Convolutional Neural Network and probability estimation of disease's occurrence.*

CNNs represent a class of artificial intelligence methods specialized for problems concerning images. The most common usage of CNNs is the classification of images in a specific number of classes. A CNN usually takes an order 3 tensor as input. For instance, a color image with M rows, N columns and 3 channels (in the RGB system) is an order-3 tensor, denoted as $X^1 \in \mathbb{R}^{M \times N \times 3}$. In binary classification problems, images represent two classes while every image takes a label depending on the class it belongs. In our study, we define that images depicting normal heart beats will be labeled with 0 and images depicting abnormal heart beats caused by arrhythmogenic right ventricular cardiomyopathy (ARVC) will be labeled with 1. A CNN consists of a series of successive layers, e.g., convolutional layers, pooling layers, batch normalization layers, fully connected layers and a loss layer. These layers have various roles in the classification process, representing the two main parts of a CNN, which are feature extraction and feature selection. The target of these deep learning algorithms is an efficient training–testing process that utilizes only the important features of our dataset's images.

### 2.3.1. Convolutional layer.

It is the most signature layer of a CNN, belonging to its first part of feature extraction. Convolution is a local operation aiming at the extraction of various patterns of the input images, resulting in an efficient classification. A CNN usually contains a series of convolutional layers improving its classification performance. Convolutional layers are consisted of multiple convolutional kernels that represent the trainable parameters of each layer. These parameters are modified during each iteration, to improve successively the pattern recognition. Let $\mathbf{X}^k \in \mathbb{R}^{M^k \times N^k \times D^k}$ be the input of the k-th convolutional layer and $\mathbf{F} \in \mathbb{R}^{m \times n \times d^k \times S}$ an order-four tensor representing the s kernels of k-th layer of spatial span $m \times n$. The output of the k-th convolutional layer will be an order three tensor denoted as $\mathbf{Y}^k$ (or $\mathbf{X}^{k+1}$) $\in \mathbb{R}^{M^k-m+1 \times N^k-n+1 \times S}$ the elements of which are the result of the convolutional operation

$$y_{i^k,j^k,s} = \sum_{i=0}^{m} \sum_{j=0}^{n} \sum_{l=0}^{d^k} F_{i,j,d^k,s} \times x^k_{i^k,j^k,l}. \tag{9}$$

Equation 2 is repeated for all $0 \leq s \leq S$ and for any spatial location satisfying $0 \leq i^k \leq M^k - m + 1$ and $0 \leq j^k \leq N^k - n + 1$.

### 2.3.2. Pooling Layer.

Let $\mathbf{X}^k \in \mathbb{R}^{M^k \times N^k \times D^k}$ be the input of the k-th layer that is now a pooling layer and its spatial extent is $m \times n$. These layers are parameter free, which means that there are no parameters that should be trained. We assume that m divides M and n divides N and the stride equals the pooling spatial extent. The output is an order-three tensor denoted as $\mathbf{Y}^k \in \mathbb{R}^{M^{k+1} \times N^{k+1} \times D^{k+1}}$, where

$$M^{k+1} = \frac{M^k}{m}, \quad N^{k+1} = \frac{N^k}{n}, \quad D^{k+1} = D^k, \tag{10}$$

while polling layer operates upon $\mathbf{X}^k$ channel by channel independently. Each $M^k \times N^k$ matrix of the order three input tensor is divided into $M^{k+1} \times N^{k+1}$ subregions where each subregion being of size $m \times n$. In most cases, the aforementioned subregions are not overlapping because nonoverlapping subregions result higher classification performances. The pooling operation maps every input's subregion into a single number $c \in \mathbb{R}$. In our CNN we used max pooling and the outputs of the pooling layers are produced according to

$$y_{i^k,j^k,d} = \max_{0 \leq i \leq m, 0 \leq j \leq n} x^k_{i^k \times m+i, j^k \times n+j, d}, \tag{11}$$

where $0 \leq i^k \leq M^k, 0 \leq j^k \leq N^k$ and $0 \leq d \leq D^k$. Intuitively, pooling layers decrease the dimensions of the output tensors while maintain the most crucial detected patterns for the purposes of the classification [56].

### 2.3.3. Fully Connected Layer.

These layers belong to the second part of a CNN and its role is the efficacious selection of features extracted from the first part of the network, targeting a high classification performance. The input of the first fully connected layer is a high dimensional vector containing all the extracted features produced by a flattening operation. A CNN often contains consecutive fully connected layers, while each layer consists of a series of nodes, representing the trainable parameters of the layer. After the last fully connected layer there is always a classification function, producing a real value $y_j$ that is compared with the expected value (label) $\hat{y}_j$ based on the selected loss function. In our case, we deem that the ideal selection is the sigmoid classification function which is defined as

$$y_j = \frac{e^{x_j}}{1 + e^{x_j}}, \qquad x_j \in \mathbb{R}, \tag{12}$$

resulting in $y_j \in (0, 1)$, intuitively representing the probability that the input image depicts an abnormal heartbeat.

Another important concept concerning the architecture of a CNN is the dropout operation, a technique used to improve the generalization of the learning method minimizing the probability of overfitting's emergence. During dropout, parameters connected to a certain percentage of nodes in the network are set to zero.

### 2.3.4. Performance Measures.

After the completion of the training phase, we must test the performance of our model using a series of validation measures. For this purpose, we will invoke three well-known measures, which are accuracy, sensitivity and specificity that can be defined [36] as

$$Accuracy = \frac{TP + TN}{TP + TN + FP + FN}, \tag{13}$$

$$Sensitivity = \frac{TP}{TP + FN}, \tag{14}$$

$$Specificity = \frac{TN}{TN + FP}. \tag{15}$$

where $TP$ represents the number of true positive instances, $TN$ of true negative, $FP$ of false positive, and $FN$ the number of false negative classified instances.

Sensitivity represents the probability that the CNN diagnoses an abnormal heartbeat given that the individual is suffering from the condition. On the other hand, specificity represents the probability that the CNN classifies the beat as normal, given that the individual is healthy. We will denote specificity (spec) as

$$P(N|H) = \frac{TN}{TN + FP} \tag{16}$$

and sensitivity (sens) as

$$P(N'|H') = \frac{TP}{TP + FN}, \tag{17}$$

where $N$ represents the fact that subject's heartbeat is classified as normal, and $H$ represents the event that the subject is healthy while $H'$ represents the existence of ARVC.

2.3.5. Probability estimation of disease's occurrence.

Our dataset contains ECG papers corresponding to patients without any heart abnormalities and patients suffering from ARVC. Each ECG paper consists of 12 signals produced from the 12 leads. The effects of arrhythmogenic cardiomyopathy affect most commonly V1 and V2 and sometimes V3 lead. As a result, we isolated for our analysis only 2 (V1 and V2) of the 12 signals from both healthy and nonhealthy patients. Some characteristic elements that accompany this specific disease are the emergence of ε waves, wider QRS complexes, inverted T waves and finally prolonged S waves [24]. Every signal of the ECG paper contains more than one beat, implying that we need an analytic expression for the probability of an individual to suffer from the specific disease, according to the number $x$ out of $n$ examined beats that are classified as abnormal, denoted as $P(H'|{}^{n}/_{x})$.

Our purpose is to display the aforementioned probability according to the sensitivity and specificity acquired from our CNN. The random event, given that the subject suffers from the disease and the CNN classified $x$ out of $n$ beats as abnormal, follows binomial distribution

$$\{{}^{n}/_{x}|H'\} \sim B(n, P(N'|H')). \tag{18}$$

As a result, the corresponding probability is

$$P({}^{n}/_{x}|H') = \binom{n}{x} P(N'|H')^{x} (1 - P(N'|H'))^{n-x}$$

$$= \binom{n}{x} (sensitivity)^{x} (1 - sensitivity)^{n-x} \tag{19}$$

Hence,

$$\{{}^{n}/_{x}|H\} \sim B(n, P(N'|H)), \tag{20}$$

while

$$P(^n/_x|H) = \binom{n}{x}P(N'|H)^x(1 - P(N'|H))^{n-x}$$

$$= \binom{n}{x}(1 - specificity)^x(specificity)^{n-x}. \qquad (21)$$

Finally, according to Bayes law of total probability, $P(H'|^n/_x)$ can be defined as

$$P(H'|^n/_x) = \frac{P(^n/_x|H')P(\text{H}')}{P(^n/_x|H')P(\text{H}') + P(^n/_x|H)P(\text{H})} = \frac{sens^x(1 - sens)^{n-x}P(\text{H}')}{sens^x(1 - sens)^{n-x}P(\text{H}') + (1 - spec)^x spec^{n-x}(1 - P(\text{H}'))} (22)$$

where $P(H')$ represents the a priori probability of an individual suffering from ARVC.

*2.4. Spectrum Analysis of Normal and Abnormal ECGs.*

Concluding our analysis, we will examine the morphological elements of the amplitude spectrums of both normal and ARVC ECG signals. Fourier analysis constitutes a method that breaks up the signal into sinusoidal waves of various frequencies. For the purposes of this attempt, we computed the amplitude spectra by taking 2 times the absolute value of the fast Fourier transform (FFT), for all frequencies up to Nyquist frequency, according to the formula [57]

$$Y_k = 2\left|\frac{1}{N}\sum_{t=0}^{N-1}y_t e^{-\frac{i2\pi kt}{N}}\right|. \qquad (23)$$

Our time series is consisted of 2.200 time points contained in the detrended V1 signals of each examinee, resulting in 77 normal and 106 abnormal amplitude spectrums. Each spectrum was normalized with the maximum amplitude as proposed in [54], boosting the robustness of the later comparisons. FFT is an algorithm constructed to estimate the Discrete Fourier Transform (DFT). The FFT returns a two-sided spectrum in complex form that can be utilized to obtain the amplitude and phase characteristics of the signal. We eliminated frequencies lower than 1 Hz, because of the unreliability of the produced results in these low frequencies due to the small length of the time series. Hence, we focus during our analysis on frequencies from 1 to 20Hz with a 0.25Hz step, because the amplitude that corresponds to greater frequencies is negligible. The usage of spectral analysis is often affiliated with filtering purposes, although there are cases where the transition from time into frequency domain may unveil important characteristics for the examined signal.

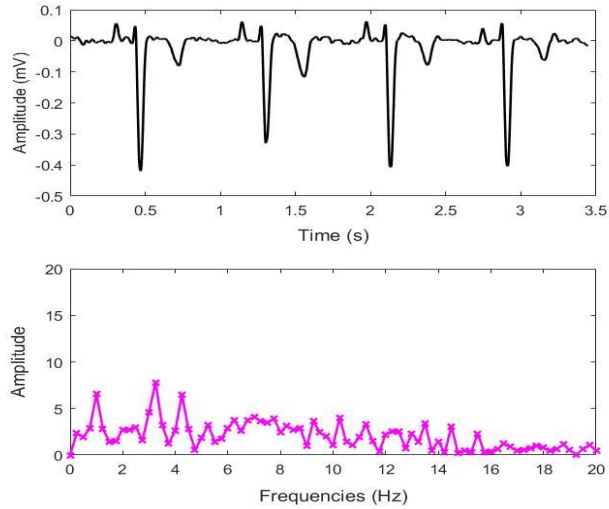

**Figure 1.** Digitalized ECG signal pertaining to V1 lead and its corresponding power spectrum.

*2.5. Dataset*

The initial dataset contains 183 paper-based ECGs, 106 of which refer to patients suffering from ARVC and 77 to healthy ones, belonging to the database of AHEPA University Hospital. The number of patients suffering from ARVC is 101, as the utilized dataset contains a second ECG for 5 of these cases. Moreover, 35 of the 77 normal ECGs correspond to male and 42 to female subjects, while 57 from the 106 ECGs that display the existence of ARVC, belong to male and 49 to female individuals respectively. The mean duration of the ECG recordings is 10 seconds. The mean age of the patients that are diagnosed with ARVC is 40.2 years with a 95% confidence interval of (33.061, 47.361), while the minimum and the maximum age of the sample is 13 and 61 years, respectively.

All abnormal ECGs that are analyzed correspond to patients with definite ARVC diagnosis. ARVC diagnosis was based on the established Task Force criteria utilizing imaging/structural criteria, ECG repolarization and depolarization abnormalities, arrhythmias family history and tissue characterization [58]. Definite diagnosis was fulfilled by the presence of 2 major, or 1 major plus 2 minor criteria or 4 minor criteria from different groups. There were no patients suffering from ARVC that had a pacemaker. From these ECGs, we emphasize the V1 and V2 leads, as it is considered that these are the leads that usually emerge the existence of ARVC.

## 3. Results

Following the aforementioned methods, in this section we will display the computational results extracted through the application of these methods to our ECG dataset.

*3.1. Validation of paper-based ECG digitalization process.*

For the digitalization process, we selected the 30 paper-based ECGs of highest quality and compared them with the digitalized signals. The validation procedure that we follow here consists of two parts. The first part includes the optical examination of ECG strips belonging to all 12 leads. In figure 2 we observe some instances of 3 digitalized ECG strips, concluding that the digitalization process successfully transmits the morphology of the paper-based ECGs to its digitalized versions, i.e., the simple digitalized and the "detrended" digitalized version.

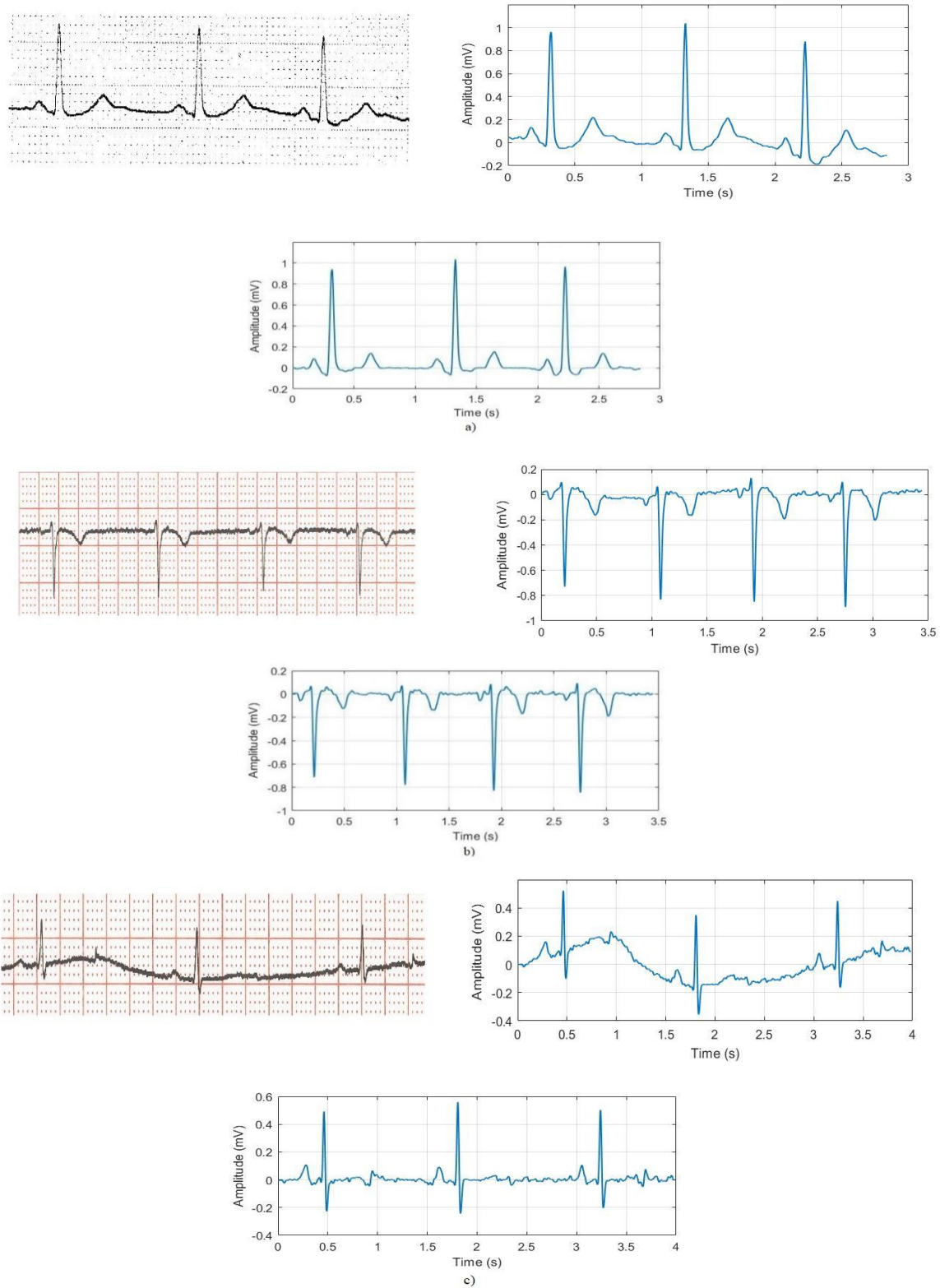

**Figure 2.** Paper ECG signals, digitalized signal and signal after detrending and smoothing from a) normal lead II, b) abnormal lead V1 and c) abnormal lead V6.

The second part of the validation procedure includes the isolation of the signal contained in the II lead of these 30 ECGs to examine the compatibility of 7 widely known ECG parameters, using Pearson correlation coefficient as shown in [24, 25, 59]. Hence, Table 1 displays the correlation coefficient, p-values and 95% confidence intervals for the duration of QRS complexes, PR, RR and QT intervals and the amplitude of P, R and T waves, while figure 3 presents the morphology of the examined parameters. The correlation values for all the intervals and amplitudes range between 0.918 and 0.995, a fact that declares a trustworthy digitalization process, while the significantly high correlation of R amplitudes and RR intervals is really promising as these quantities play an important role in the diagnosis of many arrhythmogenic cardiac diseases.

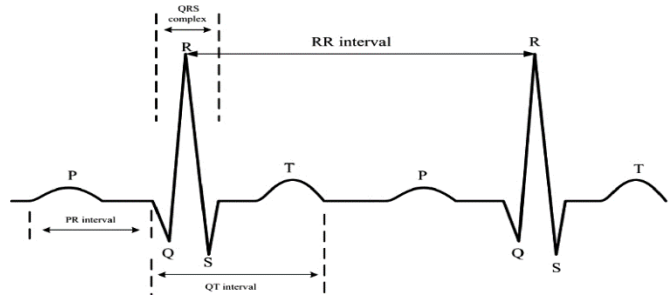

**Figure 3**. Graphical representation of the elementary ECG waves and intervals

**Table 1.** Correlation coefficients for 7 ECG characteristics between paper-based and digitalized ECG signals

| Parameter | Correlation Coefficient | p-value | 95% Confidence Interval | |
|---|---|---|---|---|
| | | | LCB | UCB |
| QRS (ms) | 0.961 | < 0.001 | 0.929 | 0.982 |
| PR (ms) | 0.944 | < 0.001 | 0.915 | 0.973 |
| RR (ms) | 0.995 | < 0.001 | 0.992 | 0.998 |
| QT (ms) | 0.963 | < 0.001 | 0.945 | 0.981 |
| P (mV) | 0.918 | < 0.001 | 0.878 | 0.958 |
| R (mV) | 0.991 | < 0.001 | 0.985 | 0.997 |
| T (mV) | 0.945 | < 0.001 | 0.917 | 0.973 |

*3.2. CNN Architecture and diagnostic accuracy.*

From the 183 ECGs contained in the dataset, we emphasize the V1 and V2 leads, as it is considered that these are the leads that usually emerge the existence of ARVC. We isolate the ECG heartbeats contained in the V1 and V2 leads of normal and abnormal patients. This procedure provides on average 8-10 ECG heartbeats from its initial ECG. The isolation of the ECG heartbeats that will be used as an input for the CNN has three main advantages in comparison of utilizing the entire patient's ECG for the training-testing process.

Firstly, this process significantly increases the number of samples consisting of the training and test set of the CNN, which seems necessary in all AI algorithms as their reliability depends highly on the size of these sets. Furthermore, there are ECGs corresponding to patients suffering from ARVC that may be distorted from premature ventricular complexes. As a result, by slicing the existing ECGs into beats we could easily exclude the distorted beats from the analysis without incapacitating the entire abnormal ECG of a disease, whose data scarcity is characteristic. Finally, as the V1 and V2 leads are split into smaller pieces, we significantly decrease the input's dimensions leading to a much faster training-testing procedure.

After applying the abovementioned procedure our dataset contains 781 images of normal and 998 images of abnormal heartbeats presenting ARVC characteristics, extracted from paper-based ECGs. Each extracted image was converted into grayscale while the underlying grid was removed, reducing noise. We used augmentation techniques to increase the amount of input images by producing slightly distorted versions of the initial heartbeat images, aiming to provide a more robust model avoiding the disadvantage of overfitting while enhancing the model's performance. Through the augmentation, we multiplied the class of normal ECG beats by 7 times and the class of the abnormal one by 6 times – aiming to produce a more balanced training-testing dataset – leading to a total of 5467 normal and 5988 ARVC beats.

**Table 2.** Structure of the proposed CNN for binary heartbeat classification.

| Layer | m x n x d | Activations |
|-------|-----------|-------------|
| Conv. | $7 \times 7 \times 32$ | ReLU |
| Batch Norm. | | |
| Conv. | $5 \times 5 \times 64$ | ReLU |
| Batch Norm. | | |
| Max Pooling | Kernel: $16 \times 16$ | Stride: 16 |
| Fully Conn. | Output Units: 1024 | ReLU |
| Dropout | Rate: 0.3 | |
| Fully Conn. | Output Units: 256 | ReLU |
| Dropout | Rate: 0.2 | |
| Fully Conn. | Output Units: 1 | Sigmoid |

In figure 4, we display the proposed CNN architecture. Our purpose is the construction of a simplistic neural network model, ideal for the confrontation of a binary classification problem characterized from low complexity input images. The training phase consisting of 100 epochs, was based on the application of the stochastic gradient descent method with a learning rate of 0.01, that was chosen due to the smooth convergence of classification loss. The following results were produced based on 50 applications of our model on different training and test sets, produced by shuffling the dataset before splitting it into these 2 sets with an 80:20 ratio, increasing the credibility of our method. Special attention is paid to the fact that heartbeats of the same patient are not included both in training and test set simultaneously. The CNN achieved 98.6% accuracy, 98.25% specificity and 98.9% sensitivity, with 95% confidence intervals being (0.9856, 0.9864), (0.9811, 0.9839) and (0.9882, 0.9898) respectively (displayed in Table 3). Figure 6 provides a visualization of the accomplished mean classification produced by the multiple applications of training-testing procedure, highlighting the trustworthiness of the proposed CNN.

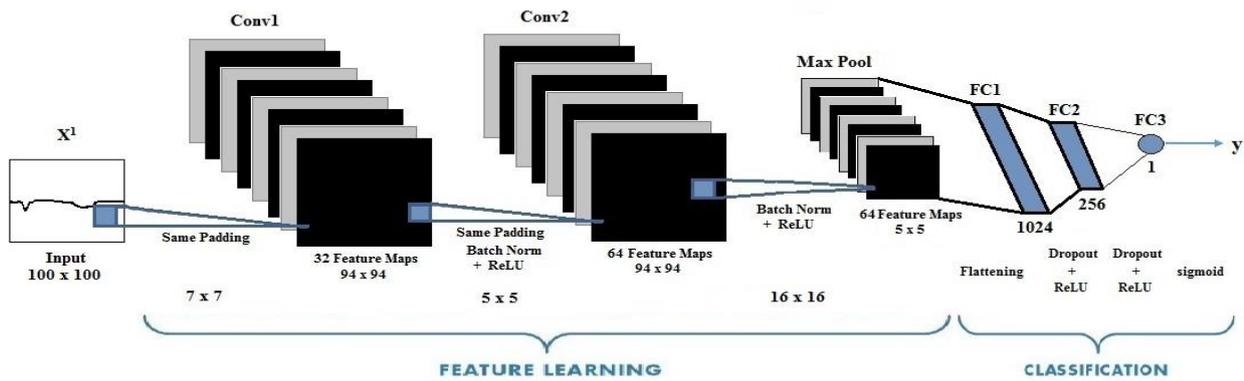

**Figure 4.** Proposed CNN architecture for the classification of normal and ARVC heartbeats.

At this point, we should emphasize that more complex architectures were examined, while they either affected negatively the classification accuracy or caused overfitting effects. We should highlight that the accomplished levels of accuracy in this article, should be deemed as really encouraging, considering that the major criteria for the identification of ARVC are infinitesimal variations in the morphology of the ECG (epsilon waves, QRS delayed S-wave upstroke with TAD $\geq$ 55ms in the right precordial leads), compared to the corresponding ECGs of healthy individuals, while the existence of these criteria is not necessary for all patients.

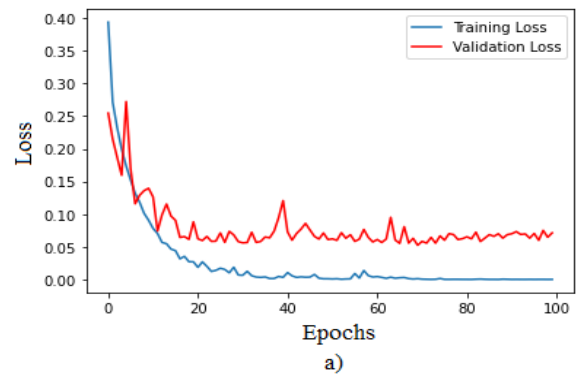

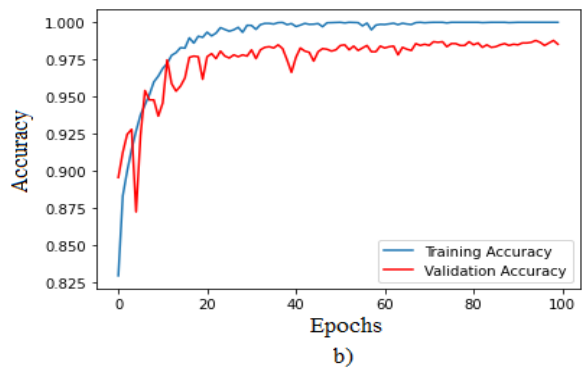

**Figure 5.** a) Training – testing loss and b) accuracy during 100 epochs.

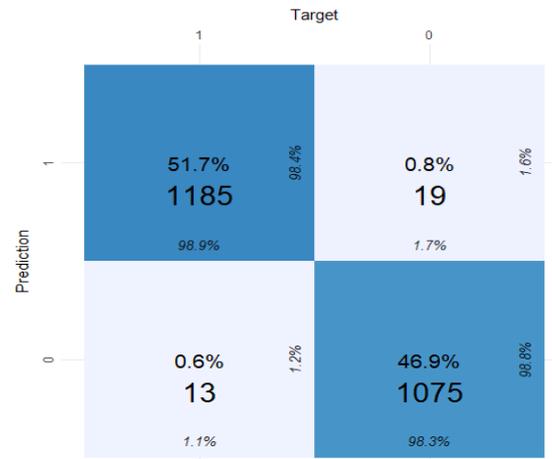

**Figure 6**. Confusion matrix of the low – complexity CNN model for the test data

**Table 3.** Table displaying the mean accuracy, specificity and sensitivity of the CNN training-testing process

|          | Accuracy      | Specificity   | Sensitivity    |
|----------|---------------|---------------|----------------|
| Training | 99.98±0.005%  | 99.96±0.01%   | 99.98±0.007%   |
| Testing  | 98.6±0.04%    | 98.25±0.14%   | 98.80±0.08%    |

As we mentioned above, we are able to calculate through equation (22) the probability that an individual is suffering from ARVC. Without any prior information for the patient's status, we assume $P(H') = 0.001$ being a mean estimation of disease's rate in the population. Considering that from an ECG paper, we may collect 10-20 heartbeats from leads V1 and V2, we display figure 7 providing the probabilities that an individual suffers from ARVC when $x$ out of $n$ heartbeats were diagnosed as abnormal.

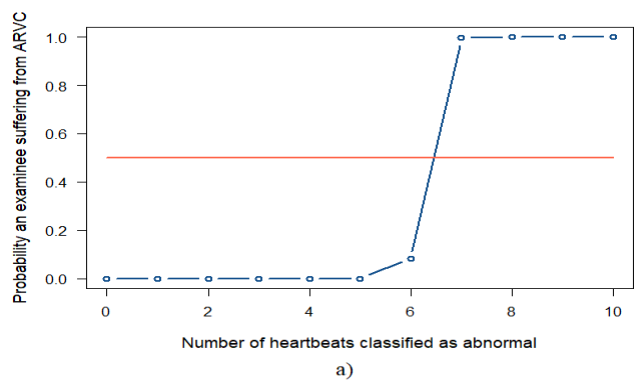

a)

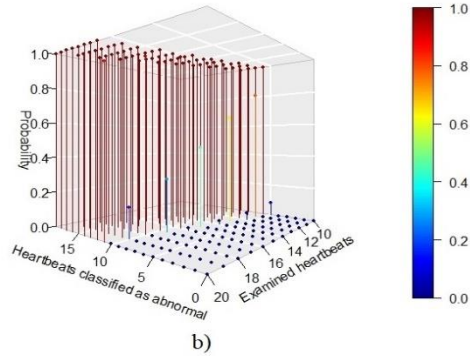

b)

**Figure 7.** a) Probability of suffering from ARVC when $x$ out of 10 ECG heartbeats are classified as abnormal, b) when $x$ out of $n$ heartbeats are classified as abnormal (n varies from 10 to 20).

*3.3. Spectral analysis.*

This paragraph is dedicated to the spectral analysis of digitalized ECGs, in order to reveal similarities and differences between normal and abnormal signals pertaining to patients suffering from ARVC. For this approach, we constructed the normalized amplitude spectrums of 77 normal and 106 abnormal digitalized ECG signals corresponding to V1 lead, utilizing the formula described in (23) and transitioning from time into frequency domain.

Our objective is the investigation of characteristics derived from the entire set of normal and abnormal digitalized ECGs. More specifically, we constructed the mean amplitude spectrums for these two categories of ECGs by calculating the mean amplitude for each frequency independently. The figure below displays the mean power spectrums of normal and abnormal ECGs.

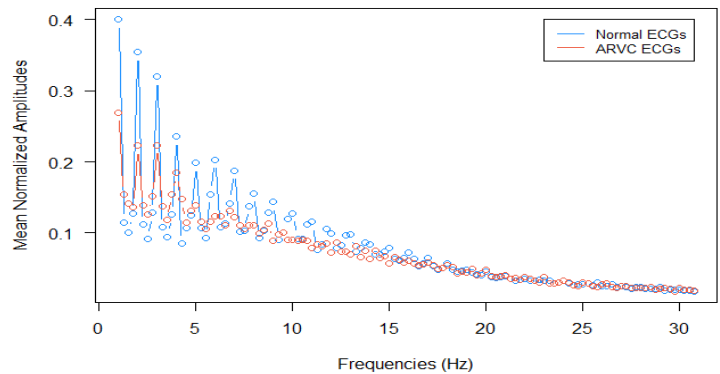

**Figure 8.** Mean power spectrums of normal ECGs and ECGs from patients suffering from ARVC.

Observing figure 8, we assume that the V1 lead signals of healthy examinees are constituted of sinusoidal waves with larger normalized amplitudes than the signals of patients suffering from ARVC. Moreover, peaks and valleys are noticed on same frequencies for both normal and abnormal ECGs, implying that in case of reconstructing signals with a finite series of sine waves, the finite sums will contain waves of the same frequencies. Another considerable observation is that in both categories, the fundamental frequency appears to be located around 1Hz, while the majority of peaks correspond to its harmonics. Finally, we could restrict our study between frequencies 1 and 20Hz, since the normalized amplitudes for more than 20Hz tend to values much smaller than 0.1.

We begin the statistical analysis by testing the normality of amplitudes' distribution, by means of Kolmogorov – Smirnov test for both normal and abnormal ECGs and frequencies 1 – 20Hz. In case of normal ECGs, the normality hypothesis cannot be rejected for 18 frequencies, 1, 2, 3, 3.75, 4, 4.25, 5, 6.75, 7.75, 9, 10, 11, 12, 13, 14, 15, 16, 17, 18 and 18.25Hz. On the other hand, referring to abnormal ECGs, the only frequencies that the normality hypothesis cannot be rejected are 1, 3 and 14Hz. As a result, we need a non-parametric statistical test to compare mean values of normalized amplitudes for all frequencies except for the 3 aforementioned cases, where normality can be assumed for both categories. Utilizing Wilcoxon – Mann – Whitney and t – test for the 3 excepted frequencies, a statistically significant difference of means arises in fundamental frequency and its 14 first harmonics plus in 3.25, 4.25 and 9.75Hz showing that the distinction between normal and abnormal ECGs should be mainly based on the amplitudes of these 15 frequencies. The assumption of homoscedasticity between samples, that is necessary for the proper application of t – test, was checked via Levene's test, where the equality of variances was satisfied only in 3Hz. Figure 9 displays graphically the distribution of p-values that has been produced by the implementation of the Mann-Whitney and T-test comparing the amplitude values of normal and abnormal ECGs per frequency.

**Table 4.** T-test for fundamental frequency, 2nd and 13rd harmonics.

| | | | | T – Test | | |
|---|---|---|---|---|---|---|
| Freq | $\overline{X}_N$ | $SD_N$ | $\overline{X}_A$ | $SD_A$ | t-statistic | p-value |
| 1Hz | .400 | .24 | .269 | .140 | 4.0472 | < .001 |
| 3Hz | .319 | .149 | .223 | .126 | 4.0249 | < .001 |
| 14Hz | .078 | .053 | .057 | .041 | 2.9011 | < .001 |

**Table 5.** Means Comparison via Wilcoxon – Mann- Whitney test.

| | | | | Wilcoxon – Mann –Whitney Test | | |
|---|---|---|---|---|---|---|
| Freq | $\overline{X}_N$ | $SD_N$ | $\overline{X}_A$ | $SD_A$ | W-statistic | p-value |
| 2Hz | .354 | .243 | .223 | .145 | 2869 | .0026 |
| 3.25Hz | .108 | .090 | .118 | .101 | 1647 | .0154 |
| 4Hz | .236 | .109 | .184 | .115 | 2759 | .0118 |
| 4.25Hz | .085 | .061 | .147 | .097 | 1245 | < .001 |
| 5Hz | .198 | .114 | .139 | .098 | 2882 | .0022 |
| 6Hz | .203 | .120 | .123 | .085 | 3115 | < .001 |
| 7Hz | .188 | .118 | .123 | .074 | 2867 | .0027 |
| 8Hz | .156 | .100 | .111 | .074 | 2790 | .0079 |
| 9Hz | .143 | .086 | .088 | .060 | 3047 | < .001 |
| 9.75Hz | .120 | .081 | .091 | .072 | 2692 | .0264 |
| 10Hz | .127 | .081 | .091 | .063 | 2813 | .0058 |
| 11Hz | .116 | .075 | .079 | .053 | 2857 | .0031 |
| 12Hz | .100 | .063 | .073 | .050 | 2745 | .0141 |
| 13Hz | .098 | .056 | .070 | .042 | 2818 | .0054 |
| 15Hz | .079 | .044 | .058 | .039 | 2819 | .0054 |

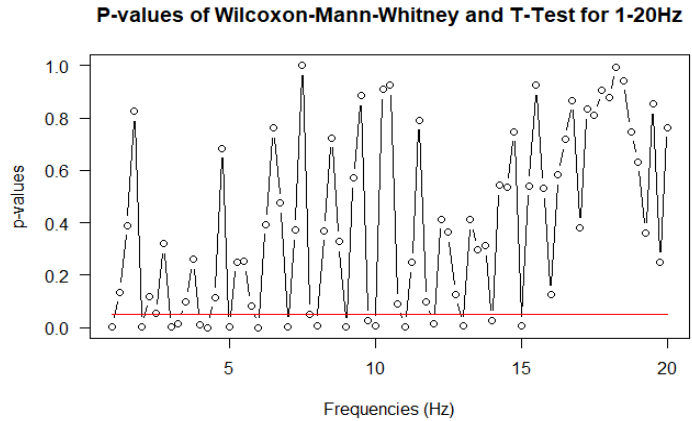

**Figure 9.** P-values of Wilcoxon-Mann-Whitney and T-test.

Mean values and standard deviations of normalized amplitudes follow a descending trend for both categories while these two statistics seem to be constantly greater in normal ECGs except frequencies 3.25 and 4.25Hz that do not belong to fundamental's frequency harmonics. In figure 10, 18 paired boxplots display the differences in the distribution of normalized amplitudes of normal and ARVC ECGs. Results show systematically lower energy of abnormal ECGs in frequencies where there is statistically significant difference, representing weaker heart voltage activity recorded by V1 lead.

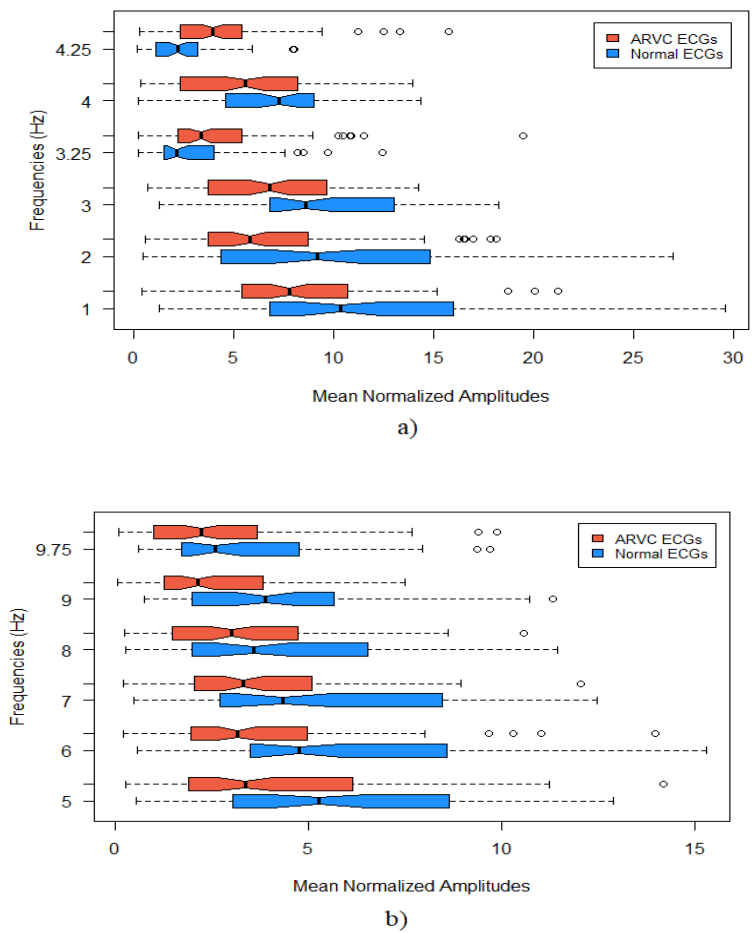

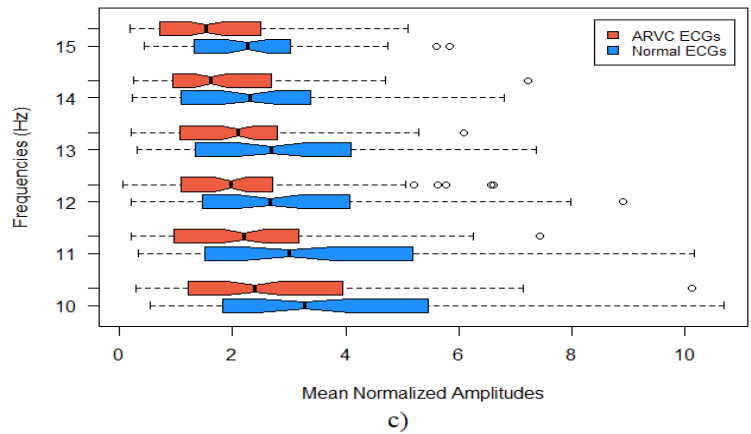

c)

**Figure 10.** Boxplots displaying a visual comparison of statistically significant differences in means of normalized amplitudes for frequencies a) $1 - 4.25$Hz b) $5 - 9.75$Hz and c) $10 - 15$Hz.

## 4. Discussion

There has been a dynamic increase in publications applying mathematical models to perform various tasks in medicine, like disease diagnosis and prediction, feature extraction and estimation of the evolution of a patient's life expectancy, aiming to improve the lifespan and living conditions of the average person. Undoubtedly, great emphasis is given in sectors like signal processing, statistics and artificial intelligence, due to the compatibility of the proposed methods in a plethora of research fields. Here, we outline the usefulness of these models in the detection and analysis of a heart disease which has not been widely investigated till now, namely ARVC, based on an everyday medical examination like ECG.

Our dataset consists of normal and abnormal ARVC ECG papers, providing the necessity of a trustworthy digitalization algorithm, since low quality digitalization could lead to disputed results. Generally, ECGs constitute a routine diagnostic tool, used from every cardiologist providing much information about probable irregularities in the heart function of each patient, especially in the hands of experienced doctors. On the other hand, the diversity in ECG's morphology due to various existing diseases and different patients can usually confuse novice doctors or practitioners culminating into wrong decisions and inefficient treatments. Moreover, modern medicine is increasingly using automated procedures based on computers and statistics, aiding cardiologist's duty and helping in deriving medical decisions in complicated cases.

For the above reasons, the requirement for credible digitalization of ECGs becomes imperative. Fortunately, both in our paper and in aforementioned articles, this requirement seems to be resolved offering the ability to cardiologist to measure signature time intervals of a heartbeat like RR, QT or PR interval, and to scientist the ability to investigate more systematically the dynamics of each heart disease, captured by ECGs liberating information trapped into paper format.

Attention have been paid in recent years to artificial intelligence methods and more importantly CNNs have received a lot of attention in recent years, because of their abundance in applications and excellent classification performances. In our work we utilized ECG signals of V1 and V2 leads as it has been proven in the literature that these are the signals where cardiac abnormalities can be observed concerning this specific disease. A common approach met in related CNN classification problems was the usage of the whole signal belonging to one or more leads as input. Although, we preferred to isolate each heartbeat that characterize the V1 and V2 signals of every ECG, producing a larger dataset satisfying neural network's necessity of numerous instances in both training and testing set. In conclusion, each heartbeat contains all the signature elements that characterize ECGs of patients suffering from ARVC, making our approach reliable and effective.

Before the training phase we augmented our dataset containing normal and abnormal heartbeats by applying transformations to the initial images, avoiding the case of creating uncommon morphologies in ECG beats. Furthermore, the proposed CNN architecture occurred after considerable experimentation aiming to construct a not complex structure, ideal for smaller datasets and binary classification problems, while eliminating the constant threat of overfitting. Moreover, the produced classification accuracy highlights the usefulness of AI algorithms that are trained according to the knowledge of experienced doctors that can base their diagnosis on additional diagnostic criteria. For instance, the ECG criteria that are considered for the diagnosis of ARVC constitute a very important tool, but they lack specificity. To lend more support, ECG is abnormal at around 90% of ARVC patients, however, none of the reported abnormalities can safely distinguish ARVC from other cardiac conditions [14]. Therefore, other task force criteria are utilized to proceed to diagnosis. On the contrary, the AI provides considerably high sensitivity, specificity and diagnostic accuracy based solely on ECG, as its training has been applied to ECG instances that were considered as abnormal, with the aid of other supplementary medical tools and not only by their visual examination.

In contrast to other image classification problems where the complexity of images is increased, the nature of our problem allows us to decrease the dimensions of the input images, significantly reducing the essential training period. Larger input dimensions or more complex CNN architectures do not improve classification accuracy while in some cases the performance is deteriorated. Moreover, as mentioned above, Bayes theorem allows us to estimate on every occasion, the probability for an individual of suffering from ARVC according to multiple classified heartbeats.

An interesting concept would be the addition of more diseases in the classification process accompanied by a multilayer CNN structure able to encounter the difficulty of multiclass classification problem. In this occasion, the calculated probabilities used in paragraph 3, should be expressed according to Bayes theorem and multinomial distributions as we move from binary to multiclass problems. The accuracy achieved should be considered promising, given the fact that the major criteria for the identification of ARVC are infinitesimal millivolt variations in the ECG's morphology, compared to other arrhythmias where the diagnostic criteria are much more prevalent, while their existence is not necessary in all individuals suffering from this cardiovascular disease.

Concerning the part of spectral analysis, we revealed a variety of hidden patterns which characterize normal and abnormal ECG signals while highlighting the important elements of frequently used tools such as the discrete Fast Fourier Transform and the Amplitude Spectrum. Our methodology emphasizes the dynamics that regulate the structure of normal V1 lead signals and compares them with the corresponding dynamics of ARVC. During this process we concluded that ARVC does not affect the mean construction of V1 signal, so both mean spectra consist of peaks and valleys at equivalent frequencies.

The fundamental frequency remains the same in both ECG categories while the majority of energy is concentrated at 1Hz and the first 15-18 harmonics. The main difference revealed by the statistical comparison, is that ARVC ECGs have lower energy in equivalent frequencies than normal, meaning a more attenuated voltage emission in the specific lead.

To summarize with, there are numerous articles that emphasize the digitalization of ECGs or the usage of CNNs for classification purposes separately. However, our article provides a complete methodology for the digitalization of scanned paper-based ECG signals and the beneficial utilization of this signals, aiming to classify robustly abnormal heartbeats corresponding to a disease that is not significantly researched through mathematical tools. The spectral analysis part of this paper provides extra characteristics for ARVC in the field of signal analysis while trying to reveal pattern differentiations between normal and ARVC ECG signals. Hence, the produced digitalization efficacy in combination with the remarkable classification accuracy and the extracted elements of the frequency domain provide new perspectives for the examination of this difficult to identify heart disorder as this is the first time that these mathematical tools have been invoked for research concerning ARVC.

## 5. Conclusions

As it is perceived, more and more scientists aim to solve problems in modern medicine using a variety of mathematical and statistical tools deploying means that have not been used for decades.

Through our analysis we proposed a complete methodology to bring usability to routine diagnostics such as ECG signals, by achieving a dynamic digitalization procedure. Our digitalization methodology provides highly accurate results as the correlation coefficients of all 7 examined ECG characteristics are greater than 0.94, while for characteristics like the duration of the RR interval and the amplitude of the R peaks we approach nearly perfect positive correlation. Accompanying the digitalization part, we promoted the need for artificial intelligence and signal processing methods in the systematic treatment of the crucial stage characterizing the diagnosis of a dangerous heart disease that has not been previously investigated on the basis of equivalent models and mathematical tools. In this specific heart disease, the analysis performed only rely on 2 out of 12 ECG leads to calculate the probability for a patient of suffering from ARVC, with significant accuracy. Namely, we propose the utilization of a low – complexity CNN that accomplishes significantly trustworthy classification performance with an accuracy of 99.98% and 98.6%, a sensitivity of 99.98% and 98.8% and a specificity of 99.96% and 98.25% for the training and testing processes, respectively. Furthermore, computationally economic CNN architectures appear to be ideal for the manipulation of medical imaging problems, as low – complexity structures can be proven as an efficacious solution against the danger of overfitting that accompanies small datasets used for diagnostic purposes.

In addition, exploring the differences of normal ECGs and ECGs belonging to patients with definite ARVC in the frequency field, we conclude that normal cardiograms are generally characterized by more intense amplitudes compared to the abnormal ones. Consequently, new signature characteristics are revealed through the spectral analysis process, thus providing more information about the nature of ECG signals. Finally, the above research plan could be followed by upcoming scientific articles targeting a better understanding of alternative heart diseases.

## Ethical Statement

### *Compliance with Ethical Standards*

*Conflict of interest*
The authors declare that they have no conflict of interest.

*Human and Animal Rights*
This article does not contain any studies with human or animal subjects performed by any of the authors.

*Informed Consent*
Informed consent does not apply as this was a retrospective review with no identifying patient information.